\documentclass[intlimits,twoside,a4paper]{article}

\usepackage{amsmath,amssymb}
\usepackage{graphicx}

\usepackage[T2A]{fontenc}
\usepackage[cp1251]{inputenc}

\usepackage[eqsecnum]{cmpj2}

\usepackage{color}

\issue{2015}{18}{1}{13002}
\doinumber{10.5488/CMP.18.13002}

\title[Role of attractive forces in determining
the equilibrium structure and dynamics of simple liquids]
{Role of attractive forces in determining
the equilibrium structure and dynamics of simple liquids}
\author[S. Toxvaerd]{S. Toxvaerd}
\address{DNRF centre  ``Glass and Time'', IMFUFA, Department of Sciences, Roskilde University, \\ Postbox 260, DK-4000 Roskilde, Denmark}

\date{Received July 31, 2014, in final form October 2, 2014}

\begin{document}

\maketitle

\begin{abstract}
Molecular Dynamics simulations of a Lennard-Jones system with different range of attraction show
that the attractive forces modify the radial distribution of the particles.  For condensed liquids
only, the forces within the the first coordination shell (FCS) are important, but for gases and moderate condensed fluids,
even the attractive forces outside the FCS play a role. The changes in the distribution caused by neglecting the attractive forces,
lead to a too high pressure.
The weak long-range attractions damp the dynamics and  the diffusion of the particles in gas-,
 super critical fluid- and in liquid states.
The values of self-diffusion coefficients (SDC) agree qualitatively with   a modified Cohen-Turnbull model.
The SDC-s  along the critical isotherm show no anomaly at the critical point in agreement with experimental data.
\pacs  05.20.Jj,  47.11.Mn, 64.70.pm, 65.20.De, 66.10.cg, 82.20.Wt
\end{abstract}

\section{Introduction}

Ever since van der Waals (vdW) \cite{vdW} in 1873 formulated his theory ``the continuity of the gaseous and liquid states'',
 the  understanding of simple fluids has been that repulsive and attractive forces give rise to separate modifications
 of the ideal-gas equation of state and to separate contributions to the free energy. The harsh, repulsive  forces reduce the
 available volume and determine the structure of the fluid and thus its configurational entropy; the weaker, longer-range
 attractive forces give rise to an energy effect that reduces pressure and energy compared to those of an ideal gas at the
 same temperature and density. With respect to the  effect of the longer-range attractive forces, van der Waals imagined
 that he could (quoting from Ref. \cite{rowwidom}) ``\ldots define an element of volume in a liquid which is small compared \ldots''
 with  ``\ldots the range of the intermolecular force, but large enough for it to contain sufficient molecules for us to assume
 that there is within it a uniform distribution of molecules of number density, $\rho=n/V$''.
The pressure $p(n/V,T)$ is in accordance with  van der Waals
\begin{equation}
    \left(p + \frac{n^2 a}{V^2}\right)\left(V-nb\right) = nRT.
\end{equation}
He  assumed that the main effect of the attractive forces originates from the molecules within this local sphere and estimated
this sphere to have a range of 3--6~{\AA} \cite{rowwidom}. For condensed liquids, the sphere can  be identified as the first coordination shell \cite{tox1}.

It is well known that the vdW equation of state reproduces the qualitative behavior of the fluid state \cite{Widom,Widom1}.
 Van der Waals' idea about the role of the repulsive and attractive forces lies behind Zwanzig's
 high-temperature expansion \cite{Zwanzig}, in which the contribution to the free energy
 from the long-range  forces is expressed in powers of the inverse temperature,
 the reference high-temperature system being a system with infinitely strong, purely repulsive forces.
 The success of this perturbation theory
(PT) was demonstrated by Longuet-Higgens and Widom \cite{lw},  Barker and Henderson \cite{bh}, and soon after by Weeks, Chandler, and Andersen (WCA) in their seminal paper ``Role of repulsive forces in determining the equilibrium structure of simple liquids'' \cite{wca}.

In PT, a fundamental problem is how to separate the strong repulsive attractions from the weaker long-range attractions. One possible separation was proposed by Barker and Henderson, who marked the separation at the distance where the potential is zero. WCA demonstrated, however, that by choosing instead the separation at the potential minimum, one obtains a much better agreement between the particle distributions of the system and the reference system. Doing so, the forces are separated into purely repulsive and purely attractive forces, and one may say that the original idea of van der Waals is here captured in its purest form. Numerous refinements of perturbation theory have appeared since then.
 An excellent summary of perturbation approaches until 1976 can be found in Barker and Henderson's classic review \cite{BHrmp}.

 Whereas the vdW equation is used  for simple systems in gas-, liquid- and coexisting liquid-gas states, the PT
theories  are normally only  used  for condensed  liquid  states. In WCA \cite{wca}, only fluid densities $\rho \geqslant 0.65$ were included in the investigation, corresponding
to the densities more than twice the critical density.
Here, we shall determine the effect of the attractions in a simple Lennard-Jones system on  the structure (section~3) and dynamics (section~4),
 as well as on gas-,  liquid- and  super-critical fluid states.

\section{ Perturbation theory for simple liquids}

In physical chemistry, the van der Waals force (or van der Waals' interaction), is the sum of the
attractive or repulsive forces between molecules (or between parts of the same molecule) other than
 those due to covalent bonds or the electrostatic interaction of ions with one another or
 with neutral molecules or charged molecules \cite{upac}. The term includes:
    force between two permanent dipoles (Keesom force),
    force between a permanent dipole and a corresponding induced dipole (Debye force) and
    force between  instantaneously induced  charge distributions (London dispersion force). The attractive London
dispersion  forces are the weakest forces. When considering ``simple liquids'', one normally refers to a system
of particles with van der Waals' interactions, and the perturbation theories are often formulated for systems
 of spherically symmetrical particles and with only the first induced dipole-dipole term in the expansion
 of the attractive London dispersion force \cite{bh,wca}.
The potential energy for such a system of $N$  spherical symmetrical particles in the volume $V$  (density $\rho=N/V$) and with
  interactions, given by a pair potential $u(r_{ij})$ between particle No. $i$ and $j$ separated the distance $r_{ij}$, is as follows:
\begin{equation}
 U=\sum_{i < j}^N u(r_{ij}),
\end{equation}
and the equation of state is obtained by the virial equation for the pressure
\begin{equation}
 p=\rho T+ p_w(\rho,T)=\rho T+ \rho w(\rho,T),
\end{equation}
where the contribution to the pressure from the intermolecular forces, $p_w(\rho,T)$ is given by
the value of the virial, $w(\rho,T)$
\begin{equation}
 w(\rho,T) =    -\frac{2 \pi}{3} \rho \int_{0}^{\infty}  g(r)\frac{\rd u(r)}{\rd r}r^3 \rd r
\end{equation}
by integration over the radial distribution $g(r)$ of the virial between particles at the distance $r$.

The strength of the force $f(r)=-\frac{\rd u(r)}{\rd r}$ can be separated into short range
force $f(r)$ for pair distances $r \leqslant \tilde{r}_\textrm{c}$ and  the long range force for  $r > \tilde{r}_\textrm{c}$.
 The  separation between short- and long range forces, given by $\tilde{r}_\textrm{c}$ is in the PT taken either to be where
 the potential $u(r)$ is zero \cite{bh} or
where the force  $f(r)$ is zero \cite{wca}.

In PT, the mean distribution of the particles,  $\tilde{g}(r)$ is obtained  by taking only forces for  $r \leqslant \tilde{r}_\textrm{c}$ into account in
the dynamics, and treating the virial contribution to the pressure for distances  $r > \tilde{r}_\textrm{c}$ as a mean field
contribution, i.e., by assuming that
\begin{equation}
\tilde{g}(r) \approx g(r).
\end{equation}
The pressure is approximately  obtained in the PT as follows:
\begin{equation}
  p(\rho, T)  \approx \tilde{p}(\rho, T) + \Delta p(\rho, T),
\end{equation}
where
\begin{equation}
\tilde{p}(\rho, T)=\rho T   -\frac{2 \pi}{3} \rho^2 \int_{0}^{ \tilde{r}_\textrm{c}} \tilde{g}(r)\frac{\rd u(r)}{\rd r}r^3 \rd r
\end{equation}
 is the pressure for a system with only short range forces for  $ r \leqslant \tilde{r}_\textrm{c}$ and the value of
the last  term
\begin{equation}
 \Delta p(\rho, T) = - \frac{2 \pi}{3} \rho^2 \int_{ \tilde{r}_\textrm{c}}^{\infty}  \tilde{g}(r)\frac{\rd u(r)}{\rd r}r^3 \rd r
\end{equation}
is  obtained as a mean field term  and using the the distribution $ \tilde{g}(r)$ for a system with only short range interactions.

The pressure (and free energy) can be derived from the exact ``$ \lambda$-expansion'' \cite{hansen}.
The thermodynamic connection between $\textit{any}$ two systems with pair potential functions, $u(r)$ and $\tilde{u}(r)$ can be obtained from the pressure of a system with  the
potential
\begin{equation}
 u(r,\lambda)= \tilde{u}(r)+\lambda \left[u(r)-\tilde{u}(r)\right],
\end{equation}
  where $u(r,\lambda)$
is equal to a reference potential $\tilde{u}(r)$
 for $\lambda=0$ and equal to $u(r)$ for  $\lambda=1$. The equilibrium   pressures for  systems with  the pair potential
$u(r,\lambda)$ and with values of $\lambda $ in the
interval $ [0,1]$ establish a reversible path between the two systems, and  the total  differences in pressure, mean energy and free energy
 are then determined  by  numerical integration over  different values of $\lambda$.

  The second step in the PT is to relate   $\tilde{p}(\rho, T)$ with the
pressure $p_0(\rho_0)$ for a system of hard spheres with a diameter $r_0$ and corresponding density $\rho_0$, i.e.,
$\tilde{p}(\rho, T) \approx p_0(\rho_0,T)$. In this article we shall investigate only the validity of the first assumption in
the PT, i.e., to what extent the structure of a simple liquid is given only by the forces $f(r)$ for $r  \leqslant \tilde{r}_\textrm{c}$.

\subsection{Separation of pair interactions into  short-  and long range interactions}

The effect of the attractive forces on the structure and dynamics is obtained for  a
system of particles with
the standard Lennard-Jones (LJ) pair potential
\begin{equation}\label{LJ}
u_{\rm LJ}(r)
\,=\,4\varepsilon  \left[(\sigma/r)^{12}-(\sigma/r)^{6}\right],
\end{equation}
and the equation of state and the diffusion are determined using Molecular Dynamics (MD) simulations. In MD, the interactions for distances
$r_\textrm{c} \gg  \tilde{r}_\textrm{c}$ are neglected. There are two standard ways to neglect these (very) long-range interactions. The usual
way is to truncate and shift the potential, whereby the force, $f_{\rm SP}(r)$, is as follows:
\begin{equation}\label{shp}
f_{\rm SP}(r)\,=\,
\begin{cases}
f_{\rm LJ}(r) & \text{if} \qquad  r<r_\textrm{c}\,, \\
0 &  \text{if} \qquad  r>r_\textrm{c}\,.
\end{cases}
\end{equation}
This is referred to as a SP cut-off because it corresponds to shifting the potential below the cut-off and putting it to zero above which it ensures a continuity of the potential at $r_\textrm{c}$ and avoids an infinite force herein.
Another and more accurate  way is to truncate and shift the  forces (SF for shifted forces) \cite{tox2} so  that the force goes continuously to zero at $r_\textrm{c}$, which is obtained by subtracting a constant term:
\begin{equation}\label{shf}
f_{\rm SF}(r)\,=\,
\begin{cases}
f_{\rm LJ}(r) - f_{\rm LJ}(r_\textrm{c}) & \text{if} \qquad  r<r_\textrm{c}\,, \\
0 &  \text{if} \qquad r>r_\textrm{c}\,.
\end{cases}
\end{equation}
This corresponds to the following modification of the potential   \cite{tox2}:
\begin{equation}\label{ushf}
u_{\rm SF}(r)\,=\,
\begin{cases}
  u_{\rm SF}(r)=u_{\rm LJ}(r) -(r-r_\textrm{c}) u'_{\rm LJ}(r_\textrm{c})-u_{\rm LJ}(r_\textrm{c}) & \text{if} \qquad  r<r_\textrm{c}\, , \\
0 &  \text{if} \qquad r>r_\textrm{c}\,.
\end{cases}
\end{equation}
 The advance of the SF methods is that one can use any value of $\tilde{r}_\textrm{c}$ or
$r_\textrm{c}$ without introducing  discontinuities into the forces. Here, we shall use SF with $r_\textrm{c}=4.5 \sigma$ as a good approximation for $r_{\infty}$.
The maximum attraction of LJ forces at $r=(26/7)^{1/6} \sigma=1.244 \sigma $ is  $f(\text{max. att.})=-2.3964 \epsilon/ \sigma$ and the attraction at $r_\textrm{c}=4.5 \sigma$ is $f=-0.000642 \epsilon/ \sigma$. The role
of the attractions is obtained for the  attractions in the interval $[\tilde{r}_\textrm{c},r_\textrm{c}]$. The longer range attractions for
 $r \geqslant 4.5 \sigma$ contribute only marginally to
 the structure and dynamics except in very diluted gases (see later).

\section{The effect of attractions on structure and pressure}

\subsection{The phase diagram of a LJ system}

\begin{figure}[!t]%Fig.1
\begin{center}
\includegraphics[height=6cm]{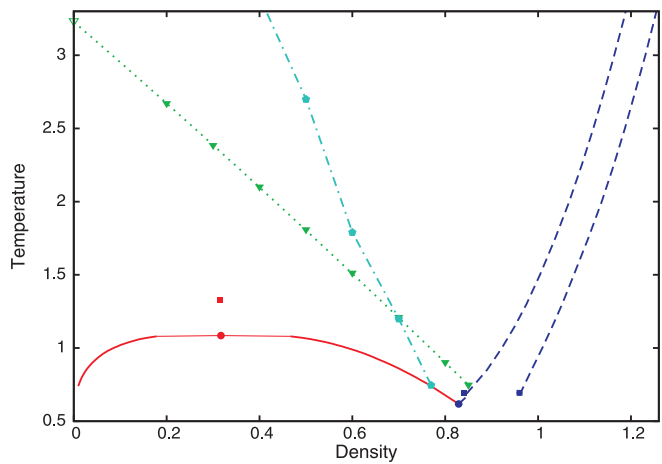}  %width=6cm, ,angle=-90
\caption{(Color online) The phase diagram $T(\rho)$ for a Lennard-Jones system. Red (solid) lines: coexisting liquid-gas \cite{Watanabe}
 with critical point for $r_\textrm{c}=2.5$ (red circle) \cite{Smit} and (red square) \cite{Caillol}  for  $r_\textrm{c}=\infty$; blue (dashed) lines coexisting  solid-liquid lines \cite{Barroso} with triple point
$(\rho_{\textrm{tr}}(l),T_{\textrm{tr}}) $ and  $(\rho_{\textrm{tr}}(s),T_{\textrm{tr}})  $ \cite{Hansen} \cite{tox3}; green (dotted) line: Zeno
line  ($\text{virial}=0$) and light blue (dash-dotted) line and points where excess energy and the virial (still) are
strongly correlated  $R=0.9$ \cite{Bailey}.}
\label{fig1}
\end{center}
\end{figure}

The attractive forces can cause a condensation of particles in a gas state into
liquid  for sufficient  low temperatures.
The phase diagram $T(\rho)$ for a LJ system is shown in figure~\ref{fig1}\footnote{Units: length in unit of $\sigma$;
time in units of $\sigma \sqrt{\epsilon/m}$; energy in units of $\epsilon/k_{\textrm{B}}$.}.
The coexisting lines and points are all obtained by various MD simulations.
There are several points to notice.
The bimodals for coexisting gas and liquid [red (solid) lines] are sensitive to the value $r_\textrm{c}$ for the
truncation of the force field. The lines in the figure are for $r_\textrm{c}=3 \sigma$ \cite{Watanabe} connected with red short dashes
to the critical point $(\rho_\textrm{c},T_\textrm{c})$, obtained by Smit \cite{Smit} for $r_\textrm{c}=2.5 \sigma$. Also shown in the figure (with a square)
 is the critical point
 $(\rho_\textrm{c},T_\textrm{c})=(0.316,1.326)$ for  $r_\textrm{c}=\infty$
obtained by Caillol \cite{Caillol}. The long range attractions behind a cut-off at $r_\textrm{c}=2.5 \sigma$ significantly
change the location of the critical point \cite{Caillol}, as well as the surface tension and interfacial structure \cite{Grosfils}.
 And for systems where the attractions are  only short-ranged,
 there is no longer a critical point and a liquid condensation \cite{Frenkel,Vliegenthart}.

The blue (dashed) lines in the figure \cite{Barroso} are the lines for  the coexisting liquid and solid
  determined by extrapolations to $r_\textrm{c}= r_{\infty}$.
The location of the  bimodals for coexisting liquid and solid (fcc) is less sensitive to the value of $r_\textrm{c}$ \cite{Mastny},
but the location of the triple point
is also sensitive to  the range of the attraction and is shifted from
 $T_\textrm{c} \approx 0.69$ for  $r_\textrm{c}= r_{\infty}$ (blue square)  \cite{Hansen} to
$T_\textrm{c}=0.618$ for  $r_\textrm{c}=2.5$ (blue circle) \cite{tox3}.
The light blue (dash-dotted) line and the points are the borderline where the excess potential energy and the virial still are strongly
correlated \cite{Bailey,Dyre4}.  The forces within the first coordination shell
are dominant and ensure a strong correlation between the potential energy and the virial for state points to the right of this line  \cite{tox1}.

\subsection{The zeno line}

Green full triangles in  figure~\ref{fig1} are points on the ``zeno line'' \cite{Ben} where the pressure
equals the pressure of an ideal gas. The points  are fitted with a second order polynomial (green line) and the
zeno line is weakly curved. All points at the zeno line are obtained for SF with $r_\textrm{c}=4.5 \sigma$. The point $(0,T_{\textrm{B}})$
with  the  Boyle temperature $T_{\textrm{B}}=3.2310$
is obtained from the expression for the second virial coefficient.
 The Boyle temperature for a LJ system ($r_\textrm{c}=\infty$) is $T_{\textrm{B}}=3.418$, and the value of the Boyle temperature
for  $r_\textrm{c}=4.5 \sigma$  is
  $T_{\textrm{B}}=3.2310$. It shows that
long-range attractive forces behind  $r_\textrm{c}=4.5 \sigma$ play a role in  (very) diluted gases.

 The virial, $w$ and the contribution $\rho w$ to the pressure  is zero
  for state points at the zeno line.
 The contribution  $p_w=\rho w$ to the pressure from the forces can be divided
into contributions from the repulsive and the attractive forces
\begin{eqnarray}
p_w(+)= \rho w(+) =  -\frac{2 \pi}{3} \rho^2 \int_{0}^{r_{\textrm{WCA}}}  g(r)\frac{\rd u(r)}{\rd r}r^3 \rd r > 0, \\
 p_w(-)= \rho w(-) =  -\frac{2 \pi}{3} \rho^2 \int_{r_{\textrm{WCA}}}^{\infty}  g(r)\frac{\rd u(r)}{\rd r}r^3 \rd r <0,
\end{eqnarray}
where $r_{\textrm{WCA}}=2^{1/6}$ for a LJ system. Hence, the zeno line is a borderline between the dominance of the repulsive and
the attractive forces on the pressure. For state points above the line, the  effect on the pressure from
 the virial of the repulsive forces is dominant.
 The structure of the fluid,
given by the radial distribution function ensures a positive value of the the virial, i.e. $p_w(+)>-p_w(-)$, and
$\textit{visa verse}$ for state points below the line.

 The central assumption in PT is that the structure is given
by the repulsive forces, and thus it is natural to investigate whether it is the case for state points along the zeno line,
where the contributions to the virial from the repulsive and the attractive forces cancel.
The LJ potential with only the first term in the induced multipole  interactions
 is the potential for the weakest London forces, and all real  (simple) van der Waals liquids
  have    stronger attractions
than a LJ system.

\subsection{The  structure of the fluid at the zeno line}

\begin{figure}[!b]%Fig.2
\begin{center}
\includegraphics[height=6cm]{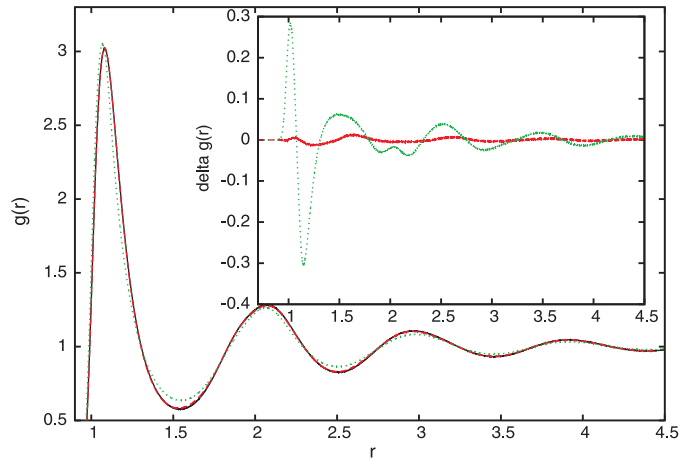}
\caption{(Color online) Radial distribution function, $g(r)$, for $\rho=0.85$ and $T=0.7453$ (at the zeno line). Black (solid): g(r);
red (dashed): $\tilde{g}(r)$ for $\tilde{r}_\textrm{c}=1.54$ (first coordination shell) and green (dotted): $\tilde{r}_\textrm{c}=2^{1/6}$ (WCA). The inset
show the difference, $\tilde{g}(r)-g(r)$. }
\label{fig2}
\end{center}
\end{figure}

\begin{figure}[!t]%Fig.3
\begin{center}
\includegraphics[height=6cm]{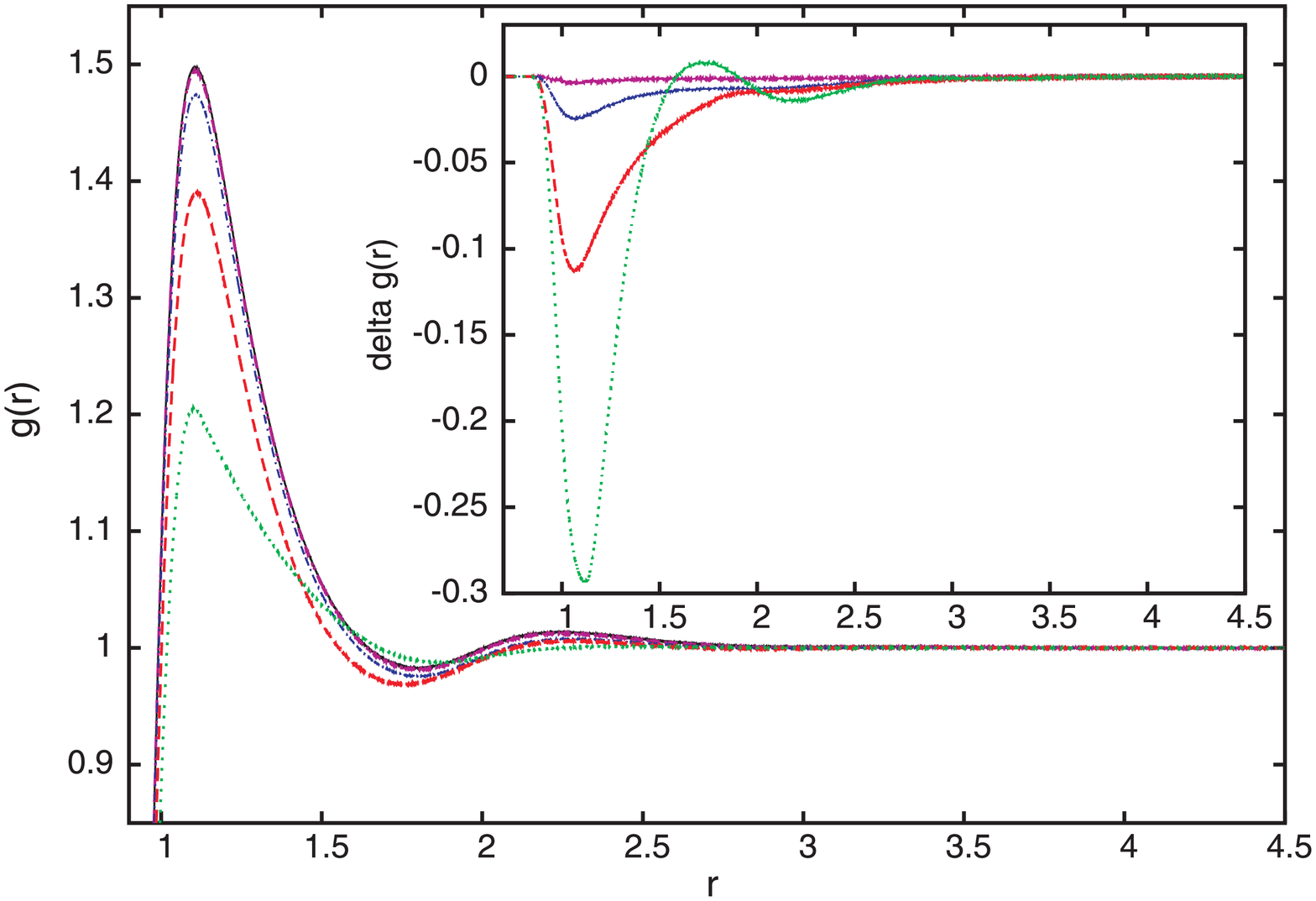}
\caption{(Color online) Radial distribution function, $g(r)$, for $\rho=0.20$ and $T=2.668$ (at the zeno line). Black (solid): g(r);
red (dashed): $\tilde{g}(r)$ for $\tilde{r}_\textrm{c}=1.75$ (first coordination shell) and green (dotted): $\tilde{r}_\textrm{c}=2^{1/6}$ (WCA). Also shown is the distribution
for  $\tilde{r}_\textrm{c}=2.5$ in blue (dash-dotted) and  $\tilde{r}_\textrm{c}=3.5$ in magenta (dash-double-dotted). The inset
show the differences, $\tilde{g}(r)-g(r)$.}
\label{fig3}
\end{center}
\end{figure}

The impact of the attractive forces on the structure of the fluid can be determined by
obtaining the structure of the fluid with and without the attraction
included in the dynamics.
 The attractive force changes the  distributions of particles. Figure~\ref{fig2} and figure~\ref{fig3} show the radial distribution functions
for the condensed liquid state  $(\rho, T)=(0.85,0.7453)$ and for the diluted gas state  $(\rho, T)=(0.20,2.6680)$, respectively.
Both state points are on the zeno line, and the effect of the attractive forces on the structure
 is bigger for the points below this line.
The black (solid) line in figure~\ref{fig2} is the  distribution $g(r)$ for a LJ fluid, and the red (dashed) line is the distribution $\tilde{g}(r)$
when  only forces within the FCS are included in the dynamics.
 The green (dotted) line is  $\tilde{g}(r)$ for $\tilde{r}_\textrm{c}=2^{1/6} $, where only
the repulsive forces are included in the dynamics. The inset shows the corresponding differences $\tilde{g}(r)-g(r)$.
 For this condensed fluid, the attractive forces behind the FCS have
only a marginal effect on the distribution, but the attractive forces within the FCS
change the structure of the fluid significantly \cite{tox1}.

 Figure~\ref{fig3}  shows the corresponding distributions at the diluted state with $\rho=0.2$. Also included [with blue (dash-dotted)]
 is the distribution
when the effect of the forces from the second coordination shell is included in the dynamics, and the magenta (dash-double-dotted) curve is for
 $\tilde{r}_\textrm{c}=3.5$.
 The two figures demonstrate  that nowhere on the zeno line there is a structure of the fluid
given only by the repulsive forces. The insets show that the main differences in the distributions
appear for distances within the FCS, and  that the differences are of the same
order for the two states. The distributions with blue (dash-dotted) and magenta (dash-double-dotted) in figure~\ref{fig3} show that whereas at condensed
states it  is the
forces within the FCS which change the structure compared to the structure of a repulsive system,
the longer range attractions behind the FCS also  have an effect on the structure of a more diluted fluid.

\subsection{Pressure  along the zeno line}

\begin{figure}[!t]%Fig.4
\begin{center}
\includegraphics[height=6cm]{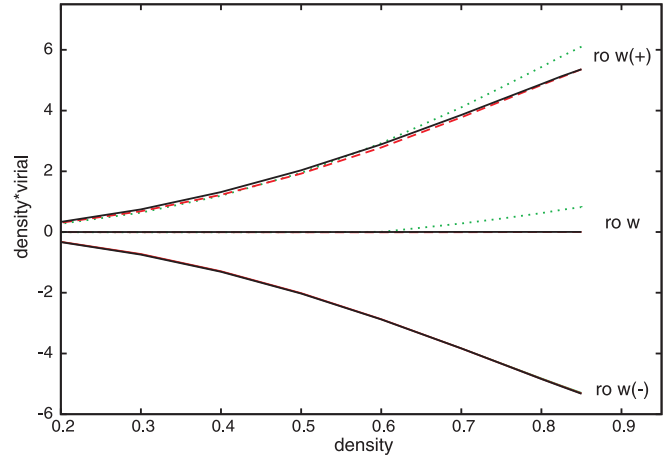}
\caption{(Color online) Contribution to the pressure from the virial, $p_w=\rho w$, along the zeno line.
 The contributions are divided into the contribution
from the repulsive forces, $\rho w(+)$ and the attractive forces,  $\rho w(-)$. With black (solid) is the two contributions and their
sum (which is equal to zero), and with red (dashed) are the corresponding contributions obtained from  $\tilde{g}(r)$ by neglecting
the attractive forces behind the first coordination shell in the dynamics. Green (dotted) are obtained from  $\tilde{g}(r)$  for a system of
repulsive LJ particles.}
\label{fig4}
\end{center}
\end{figure}

According to figure~\ref{fig2} and  figure~\ref{fig3}, the radial distribution $\tilde{g}(r)$
differs from $g(r)$ if the attractive forces are neglected in the dynamics. The integral effect on the pressure is given by
\begin{eqnarray}
\tilde{p}_w(+)= \rho \tilde{w}(+) =  -\frac{2 \pi}{3} \rho^2 \int_{0}^{r_{\textrm{WCA}}} \tilde{g}(r)\frac{\rd u(r)}{\rd r}r^3 \rd r, \\
\tilde{p}_w(-)= \rho \tilde{w}(-) =  -\frac{2 \pi}{3} \rho^2 \int_{r_{\textrm{WCA}}}^{\infty}   \tilde{g}(r)\frac{\rd u(r)}{\rd r}r^3 \rd r.
\end{eqnarray}
The  two contributions and their sum are shown in figure~\ref{fig4}. With black (solid) is $p_w, p_w(+)$ and  $p_w(-)$ for a LJ fluid,
with red (dashed) $\tilde{p}_w, \tilde{p}_w(+)$ and  $ \tilde{p}_w(-)$ for  $\tilde{g}(r)$ with forces only from the FCS.
The green (dotted) curves are  $\tilde{p}_w, \tilde{p}_w(+)$ and  $ \tilde{p}_w(-)$ for a system with only repulsive LJ forces.
Distributions are changed by neglecting the  attractive forces in the dynamics, but the integral effect of the attractions
 on the pressure,
$\tilde{p}_w(-)$ is close to zero. This is, however, not the case for the effect on the pressure
$ \tilde{p}_w(+)$ from
 changes in the distribution
for distances where the forces are repulsive.  Figure~\ref{fig4}  reveals that the pressure obtained by PT fails, not at diluted supercritical states,
but for condensed states $\rho > 0.6$, and it fails because the structure of the fluid for
distances where the forces are repulsive is changed due to the
effect of the attractive forces! Figure~\ref{fig2} shows [green (dotted) lines] that the radial distribution function $\tilde{g}(r)$ for
$r<2^{1/6}$ is ``parallel shifted'' toward shorter distances by neglecting the attraction in the dynamics, and it
is this effect which leads to the deviation   in the virial and pressure  $ \tilde{p}_w(+)-p_w(+)$ for a condensed fluid.

The apparent agreement between the radial distribution function of a LJ system and the corresponding distribution
for a system with only repulsive forces for a  liquid is striking (figure~\ref{fig2}); but there are systematic
differences at short distances and these differences lead to a too high pressure.
The standard perturbation theory (PT) (Barker-Henderson \cite{bh} and Weeks-Chandler-Anderson \cite{wca}) is not
quantitatively correct.
The central assumption in PT is that the effect of the attraction can be obtained from a fluid  with only repulsive forces.
It is, however, possible to modify the PT by replacing the reference system with a reference system with the forces
within the first coordination shell \cite{tox1}. Doing so, the pressure obtained by PT
 is  not only qualitatively and to a high extent also
 quantitatively correct for
diluted gas and super critical states as well as for condensed liquid states.

 The standard perturbation theory
is the ``zero'' term in an exact  $\lambda$ expansion \cite{hansen}, and it is, of course, possible to obtain
the pressure from  the  pressure  $\tilde{p}(\rho,T)$ for the short range potential  by determining the distribution
for the potential $u(r,\lambda)$ for sufficient values of $\lambda$  \cite{weeks,rull}.

\section{The role of  attractions on self-diffusion}

\begin{figure}[!t]%Fig.5
\begin{center}
\includegraphics[height=6cm]{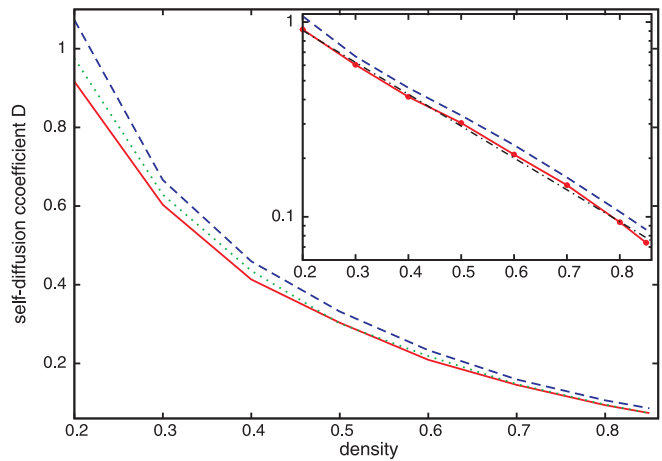}
\caption{(Color online) Self-diffusion coefficients $D(T,\rho)$ for the critical LJ isotherm $T=T_\textrm{c}=1.326$. Red (solid) line: LJ fluid; green (dotted): for a system with
only forces within the FCS included in the dynamics, and  blue (dashed) is when only the repulsive forces are included in the dynamics.
The inset shows  $\ln D(T,\rho)$ and with black (dash-dotted): $a \exp(-b \rho)$.}
\label{fig5}
\end{center}
\end{figure}

 The attractive forces change the dynamics of a system. The  effect of  the  attractive forces on
the self-diffusion coefficient (SDC) is determined for the critical isotherm $T=T_\textrm{c}$ and the
triple point fluid isochore $\rho_{\textrm{tr}}$ for a LJ system.  The  values of the SDC for
the critical LJ isotherm  $T_\textrm{c}(\textrm{LJ})=1.326$ are shown in figure~\ref{fig5}.
The values  are obtained from the mean displacements of the $N=2000$ particles and for $1.6 \times 10^6$
time steps with $\delta t=0.0025$. For the system  with only short-range attractions
[green (dotted) line], the temperature is
well above its critical temperature, and  the system with no attractions   [blue (dashed) line] has no critical temperature.
 The  SDC for all three systems decline
exponentially with the density in the interval $\rho \in [0.2,0.8]$ (inset). For  densities
$\rho>0.5$, there are no significant differences between the coefficients for the LJ system [red (solid) line] and the coefficients
for the system with only forces within the FCS in the dynamics [green (dotted) line]  \cite{tox2}, but for smaller densities also the
longer ranged forces play a role and damp the diffusion.
The diffusion in a LJ system is systematically
smaller than the diffusion for a system with only repulsive forces [red (solid) and blue (dashed) lines],
and the differences increase with decreasing density.

There is no observable change in the behavior of the SDC  of the LJ system near the critical point.
Only the LJ system exhibits critical fluctuations. The system with only short-range attractions from the FCS
has a lower critical temperature, and the system with only repulsive forces has
no critical temperature. Nevertheless, the three systems exhibit a coherent behavior and  with no spur of a critical effect,
which should be in
only  one of the three systems.
The critical point marks the location of a thermodynamic instability and the effect on the SDC is small, if any,
 and in agreement with experimental data
\cite{Drozdov,Drozdov2,Harris}.

The SDC varies  exponentially with increasing density. The inset shows  $\log D$ for the LJ system [red (solid)] and the system
without attractive forces [blue (dashed)]. The  coefficients $a$ and $b$ in
 the function $a \re^{-b \rho}$ are
fitted to the values of $D$ for the LJ system, and the function is shown in the inset [black (dash-dotted) straight line].
The data show that the role of the attraction is given by the value of the preexponential coefficient $a$, whereas
the values of the coefficient $b$ for the two systems are the same within the accuracy of the MD simulation.

The SDC are usually compared with the coefficients obtained
using a modified Cohen-Turnbull theory \cite{eu,iran}. In this theory, the diffusion coefficients are given by
\begin{equation}
 D=D_0 \exp\left ( -\alpha \frac{v_\textrm{c}}{v_\textrm{f}} \right),
\end{equation}
where $D_0$ is the Chapman-Enskog diffusion for a
model pair potential at low density
\begin{equation}
 D_0=D_0^{\textrm{HS}}/\Omega^{(1,1)}(T).
\end{equation}
 The coefficient $D_0^{\textrm{HS}}$ is the SDC for a hard sphere system,
the temperature dependent  collision integral, $\Omega^{(1,1)}(T)$,
  depends on the attractive forces \cite{Hirsfelder},
$ \alpha$ is an (``overlap'') parameter, $v_\textrm{c}$ is ``the critical volume'' and $v_\textrm{f}$ is the ``mean free volume''. The data
in figure~\ref{fig5} agree with this model with $a =D_0$, $ b = \alpha v_\textrm{c} $ and $v_\textrm{f} \approx \rho^{-1}$.

\begin{figure}[!t]%Fig.6
\begin{center}
\includegraphics[height=6cm]{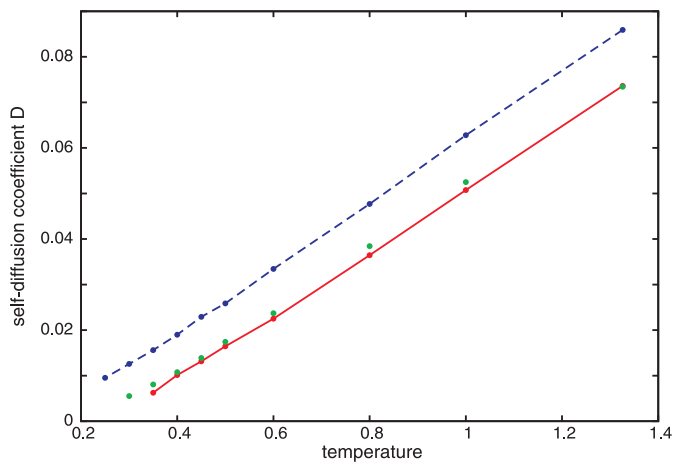}
\caption{(Color online) Self-diffusion coefficients $D(T,\rho)$ for the triple point fluid density $\rho_\textrm{tr}=0.85$. Red (solid) line and points: LJ fluid; green points: for a system with
only forces within the FCS included in the dynamics, and  blue (dashed) line and points are when only the repulsive forces are included in the dynamics.}
\label{fig6}
\end{center}
\end{figure}

The SDC for the   triple point (fluid) density $\rho_\textrm{tr}=0.85$ are shown in figure~\ref{fig6}. With red (solid) line
and points are $D$ for the LJ system. The triple point temperature is $T_\textrm{tr} =0.69$ and the system is supercooled down
below $T_\textrm{tr}$
to $T=0.30$ where it crystallizes. The diffusion coefficients for the  system with only short-range attraction (green) are
almost equal to the diffusion coefficients for the LJ system, showing that the long range attractions behind the FCS only have
a marginal effect on the diffusion at high temperatures as well as on the diffusion at super-cooled temperatures. The diffusion
coefficients for the system without attractive forces [blue (dashed) line and points] are, however, systematically higher. The
SDC for the three systems are all given by a simple linear dependence $a+b T$ where $b \approx $ independent of
the range of the attractive forces, and  $a$'s dependence on the attractive forces to a high accuracy is given by the short range forces
within the FCS. At a deeper super cooling for a system which is more
resistant against crystallization \cite{ka,ka2},  the dynamics and the SDC  can still be
obtained  with only attractive forces within the FCS included in the dynamics \cite{tox1}, whereas the SDC for a
system with only repulsive forces  deviates significantly \cite{t,tox1}.

\section{Conclusion}

Liquids are often  separated into simple van der Waals liquids with weak intermolecular  attractive forces   and into
liquids with stronger attractions.
The van der Waals theory for simple liquids, where the repulsive and the weak attractive  forces
 give rise to separate modifications of volume and pressure in the
ideal-gas equation of state, is qualitatively correct and has played a historical role in understanding the
fluid systems.   The perturbation theories (PT) for simple liquids, which were formulated late in the twentieth century,
showed good agreements with experimental-  and computer simulated data,
 and gave reason to believe that it is indeed possible
to obtain  the contribution to the pressure from the weak  attractive  van der Waals forces as a mean field
contribution  from  a system with the particle distribution for a system with  only   strong repulsive forces.
 Since then, the development in computer simulations
has made it possible to obtain the equation of state for systems with different strength of attractions with high accuracy,
and to investigate various approximations in the PT theories for fluid systems.
The present investigation shows that, even for a system with the most weak attractive London forces,
 nowhere in the phase space, it is possible to approximate the radial distribution of particles with  the distribution obtained for a system with only  repulsive forces.
Even weak attractive induced dipole forces, primarily within the first coordination shell, modify the radial  distribution, especially at
short intermolecular distances (figure~\ref{fig2}), and with the result that the pressure obtained by the PT is too high (figure~\ref{fig4}).

The weak long-range attractions damp the dynamics and  the diffusion (figure~\ref{fig5} and figure~\ref{fig6}).
Since it is the  attractive forces which cause a condensation into a liquid for temperatures below the critical temperature, $T_\textrm{c}$,
one should \emph{a priory} assume that the attractive forces, which cause  long range fluctuations in densities
at $T_\textrm{c}$ would affect the particle diffusion and the value of the   self-diffusion constant (SDC) near the critical point.
 However, this is not the case,  and the
 result is in accordance with experimental data for the SDC in the critical region \cite{Harris}.

The SDC along the critical isotherm exponentially varies with the density (figure~\ref{fig5}). This is in good agreement
 with a modified Cohen-Turnbull model for the SDC \cite{eu,iran}. The effect of the attractive
 forces on the SDC is a simple damping of the diffusion and
with a reduced value of the preexponential constant. The attractive forces also  damp the diffusion in condensed  super-cooled  fluids.
The temperature dependence of the SDC at the triple-point isochore shows a simple
 linear damping with respect to temperature, and the damping of the diffusion is caused by the attractive forces
within the first coordination shell (figure~\ref{fig6}). For systems which are less prone to crystallization and at a deeper super-cooling, the
 effect  of the attractive forces is more dominant, but the attractive forces
 within the FCS still determine the diffusion and viscosity, as well as the thermodynamics \cite{tox1}.

\section*{Acknowledgements}%
 The centre for viscous liquid dynamics ``Glass and Time'' is sponsored by the Danish National Research Foundation (DNRF).

\clearpage

\ukrainianpart

\title{Роль сил притягання у визначенні рівноважної структури і динаміки простих рідин}
 \author{С. Токсваерд}
 \address{Факультет природничих наук, Університет м. Роскільде, DK-4000 Роскільде, Данія}

 \makeukrtitle

 \begin{abstract}
 Моделювання методом молекулярної динаміки системи Лєннарда-Джонса з різною областю притягання показують,
 що сили притягання видозміюють радіальний розподіл частинок. Сили в межах першої координаційної сфери (ПКС)
 є важливими лише для густих рідин, в той час, як для газів і помірно густих плинів, навіть сили притягання
 за межами ПКС відіграють певну роль. Зміни у розподілі, спричинені нехтуванням силами притягання, приводять
 до занадто високого тиску. Слабке далекосяжне притягання гасить динаміку і дифузію частинок у станах газу,
 надкритичного плину та рідини. Значення коефіцієнтів самодифузії якісно узгоджуються з модифікованою моделлю
 Коена-Турнбулла. Коефіцієнти самодифузії вздовж критичної ізотерми не показують ніякої аномалії у критичній
 точці, що узгоджується з експериментальними даними.
 \keywords прості рідини, рівняння стану, теорія збурень, самодифузія
\end{abstract}

\end{document}